\documentstyle[12pt]{article}
\begin{document}
\begin{flushright}
hep-ph/0104167
\end{flushright}
\vspace{1 cm}
\begin{center}
\baselineskip=16pt

{\Large\bf  Nonequilibrium Temperature for Open Boson and Fermion
Systems}

\vskip 2cm {\bf Sang Pyo Kim}\footnote{E-mail:
spkim@phys.ualberta.ca}
\\ \vskip 0.8cm

Theoretical Physics Institute\\Department of Physics\\University
of Alberta\\Edmonton, Alberta, Canada, T6J 2J1\\ and \\

Department of Physics\\Kunsan National University\\Kunsan 573-701,
Korea\\

\vskip 1 cm
\end{center}

\date{\today}

\vskip 1 cm

\centerline{\bf ABSTRACT}
\begin{quotation}
The effective theory of an open boson or fermion system is
studied, which evolves out of equilibrium with a time-dependent
Hamiltonian $\hat{H} (t)$. A measure of nonequilibrium temperature
for the open system evolving from an equilibrium is proposed as
the time-averaged energy expectation value $T (t) = T_i
(\overline{\langle \hat{H} (t) \rangle_{\Psi}}/ \overline{\langle
\hat{I} (t) \rangle_{\Psi}})$, where $\hat{I} (t)$, the action
operator, satisfies the quantum Liouville-von Neumann equation and
determines the true density operator. It recovers the result for
the adiabatic (quasi-equilibrium) and nonadiabatic
(nonequilibrium) evolution from one static Hamiltonian to another
and takes into account the particle production due to the
intermediate processes.\\

\end{quotation}

\newpage

\section{Introduction}

A system becomes nonequilibrium (out of equilibrium) when its
coupling constants change rapidly through the interaction with an
environment. In inflationary scenarios, the inflation of the
Universe would rapidly cool the inflaton and matter fields and
this supercooling process would have happened far from equilibrium
(see ref. \cite{lin}). Another phenomenon out of equilibrium is
the recently introduced preheating mechanism, in which the
parametric resonance of bosons and fermions coupled to the
oscillating inflaton leads to catastrophic production of particles
\cite{bra1,bra2,kof1,tka,kof2}. In quark-gluon plasma the second
order phase transition out of equilibrium is expected to play an
important role in quark-confinement \cite{raj1,raj2,raj3}.
Recently the nonequilibrium phase transitions have been actively
studied to understand the dynamical process of domain growth and
topological defects (see ref. \cite{bun}).

The quantum statistics, however, has not attracted much attention
for these nonequilibrium systems though it is important in
understanding the multi-particle phenomenon. Further a measure of
temperature has not been assigned to such a nonequilibrium system
because temperature is widely believed to make a physical sense
only for an equilibrium system. But a quantum statistical theory
has been proposed for nonequilibrium system based on the
assumption that the microscopic world is correctly described by
the Schr\"{o}dinger equation and the quantum Liouville-von Neumann
equation \cite{kim-lee}. For the nonequilibrium system the
frequently-used operator $\hat{\rho}_{H} (t) = e^{- \hat{H} (t)/
kT}/Z_{H}$ cannot be used since it is {\it not} the true density
operator. Instead the true density operator that satisfies the
quantum Liouville-von Neumann is given by another operator
$\hat{\rho}_{I} (t) = e^{- \hat{I} (t)/kT}/Z_{I}$, from which
quantum statistics such as the Bose-Einstein or the Fermi-Dirac
distribution is derived. Hence it may be suggested that the
nonequilibrium temperature should be defined out of the
Liouville-von Neumann operators $\hat{I} (t)$ and the dynamical
energy operator $\hat{H} (t)$.

In this paper a quantitative measure of temperature is proposed
for the nonequilibrium system. The analysis will be confined to
the effective theory of an open boson or fermion system whose
Hamiltonian $H(t)$ explicitly depends on time. The spatial
homogeneity of the system will be assumed and the interaction
among constituent particles will be neglected. The time-dependent
nature of the system comes mainly from the interaction with an
environment. Further, to simplify the analysis, we focus on the
open boson or fermion oscillator with time-dependent frequency
$\omega (t)$. The time-dependent boson oscillator has been
extensively studied as an exactly solvable, nonstationary, quantum
system \cite{d-osc1,d-osc2,d-osc3}. Also an open fermion system
has been studied as the nonequilibrium system \cite{kim1}. For the
time-dependent oscillator we derive the nonequilibrium temperature
using the exact density operator $\hat{\rho}_{I} (t)$ and the
Hamiltonian operator $\hat{H} (t)$ as giving the dynamical energy.
Many solvable models are investigated, which evolve from one
static Hamiltonian to another and have the limit of the adiabatic
(quasi-equilibrium) or the nonadiabatic (nonequilibrium)
evolution.

The organization of this paper is as follows. In Sec. II the
relation between temperature and energy expectation value in the
high temperature limit is discussed qualitatively and the
temperature relation is derived approximately for the adiabatic
and nonadiabatic evolution. In Sec. III, using the action operator
and the exact density operator, the nonequilibrium temperature is
defined by the ratio of the time-averaged energy expectation value
to the expectation value of the action operator. The
nonequilibrium temperature is shown to be the quantum analog of
the classical one from classical distribution function.  In Sec.
IV the nonequilibrium temperature is applied to several models of
physical interest and extended to the time-dependent fermion
system. In the nonadiabatic evolution from one static Hamiltonian
to another, the final temperature is obtained in terms of the
initial temperature, initial and final frequencies, and the
particle production rate.

\section{Adiabatic vs. Nonadiabatic Evolution}

We start with an ensemble of boson oscillators with the frequency
$\omega$ in a thermal equilibrium at temperature $T$. The thermal
state of the system is described by the density operator
$\hat{\rho}_{H} = e^{- \hat{H} / kT}/Z_{H}$ and has the energy
expectation value
\begin{equation}
\langle \hat{H} \rangle_{\rho_{H}} = \frac{\hbar \omega}{2} \coth
\Biggl(\frac{\hbar \omega}{2 k T} \Biggr),
\end{equation}
where $\hbar$ is the Planck constant and $k$ the Boltzmann
constant. In the high temperature limit $k T \gg \hbar \omega$,
the thermal energy expectation value has the limiting value
\begin{equation}
\langle \hat{H} \rangle_{\rho_{H}} = k T.
\end{equation}
This may suggest that the temperature of the system should be
derived from the energy expectation value at least in the high
temperature limit.

We now evolve the system from one static Hamiltonian $\hat{H}_i$
to another $\hat{H}_f$ by adiabatically changing the frequency
from $\omega_i$ to $\omega_f$. If the system was in a thermal
equilibrium, it would maintain quasi-equilibrium at each moment
from the initial equilibrium to the final one. Both the initial
and final states are assumed to have high temperatures. Then the
ratio of the initial and final temperatures is given by the ratio
of the corresponding energy expectation values:
\begin{equation}
\frac{T_f}{T_i} = \frac{\langle \hat{H}_f
\rangle_{\rho_{f}}}{\langle \hat{H}_i \rangle_{\rho_{i}}},
\label{ad tem}
\end{equation}
where $\rho_i$ and $\rho_f$ denote the initial and final density
operators $\hat{\rho}_{H}$. If the density operator
$\hat{\rho}_{H}$ does not change significantly, eq. (\ref{ad tem})
can be approximated by
\begin{equation}
\frac{T_f}{T_i} \simeq \frac{\langle \hat{H}_f
\rangle_{\rho_{i}}}{\langle \hat{H}_i \rangle_{\rho_{i}}} =
\frac{\omega_f}{\omega_i}. \label{app ad tem}
\end{equation}
The particle production due to the change of frequency (parameter)
is negligible during the adiabatic evolution.

In the nonadiabatic evolution, the rapid change of frequency
drives the system out of thermal equilibrium. Such a nonadiabatic
process necessarily involves the particle production
\cite{par1,par2,dew}. The creation and annihilation operators at
the initial and final times are related with each other through
the Bogoliubov transformation
\begin{equation}
\hat{a} (t_f) = \alpha  \hat{a} (t_i) + \beta \hat{a}^{\dagger}
(t_i), \quad \hat{a}^{\dagger} (t_f) = \alpha^*  \hat{a}^{\dagger}
(t_i) + \beta^* \hat{a} (t_i), \label{bog tran}
\end{equation}
where the coefficients satisfy
\begin{equation}
\alpha^* \alpha  - \beta^* \beta  = 1. \label{bc}
\end{equation}
It is interesting to compare the expectation values of the final
and initial Hamiltonians with  respect to the number states,
$\vert n, t_i \rangle$, of initial time $t_i$, whose ratio is
given by
\begin{equation}
\frac{\langle \hat{H}_f \rangle_{n_i}}{\langle \hat{H}_i
\rangle_{n_i}} = \Biggl( \frac{\omega_f}{\omega_i} \Biggr) \times
(1 + 2 \beta^* \beta).
\end{equation}
It therefore follows that
\begin{equation}
\frac{\langle \hat{H}_f \rangle_{\rho_{i}}}{\langle \hat{H}_i
\rangle_{\rho_{i}}} = \Biggl( \frac{\omega_f}{\omega_i} \Biggr)
\times (1 + 2 \beta^* \beta). \label{nonad rat}
\end{equation}
The factor $\beta^* \beta$ is the number of particles created per
unit volume and depends on the intermediate nonadiabatic process
during the evolution. As will be shown later, eq. (\ref{nonad
rat}) is related to the temperature ratio. If a device measures
the temperature of the system with time-averaged energy density,
the final temperature would be given by
\begin{equation}
T_f =  T_i \times \Biggl[\frac{\omega_f}{\omega_i} (1
+ 2 \beta^* \beta) \Biggr]. \label{nonad tem}
\end{equation}

\section{Nonequilibrium Temperature}

To define a quantitative measure of temperature that recovers eqs.
(\ref{app ad tem}) and (\ref{nonad tem}) in the limiting cases, we
study an open boson system described by a time-dependent
oscillator
\begin{equation}
H (t) = \frac{p^2}{2} + \frac{\omega^2 (t)}{2} q^2,
\end{equation}
where the frequency $\omega(t)$ explicitly depends on time. The
quantum states of the system are governed by the time-dependent
Schr\"{o}dinger equation
\begin{equation}
i \hbar \frac{\partial \Psi(q, t)}{\partial t} =
\hat{H} (t) \Psi (q, t). \label{sch eq}
\end{equation}
The statistical properties of the system are determined by the
density operator that satisfies the quantum Liouville-von Neumann
equation
\begin{equation}
i \hbar \frac{\partial \hat{\rho} (t)}{\partial t} + [
\hat{\rho} (t), \hat{H} (t)] = 0. \label{ln eq}
\end{equation}
The physical quantities which can be directly determined by eqs.
(\ref{sch eq}) and (\ref{ln eq}) are $\langle \hat{H} (t)
\rangle_{\Psi}$ and $\langle \hat{\rho} (t) \rangle_{\Psi}$, from
which all other statistical variables such as the temperature and
free energy will be derived. Here $\vert \Psi \rangle$ denotes any
exact quantum state of eq. (\ref{sch eq}).

To find the exact quantum state of eq. (\ref{sch eq}), we use the
idea of invariant operators first introduced by Lewis and
Riesenfeld \cite{lew}, who found a quadratic invariant of position
and momentum. In refs. \cite{kim2,kim-page,kim-lee,d-osc2}, a
pairs of linear invariant operators satisfying eq. (\ref{ln eq})
are introduced
\begin{eqnarray}
\hat{a}^{\dagger} (t) &=& - \frac{i}{\sqrt{\hbar}} [\varphi (t)
\hat{p} - \dot{\varphi} (t) \hat{q}] \nonumber\\ \hat{a} (t) &=&
\frac{i}{\sqrt{\hbar}} [\varphi^* (t) \hat{p} - \dot{\varphi}^*
(t) \hat{q}],
\end{eqnarray}
where $\varphi$ is a complex solution to the classical
equation
\begin{equation}
\ddot{\varphi} (t) + \omega^2 (t) \varphi (t) = 0. \label{cl eq}
\end{equation}
The Wronskian condition
\begin{equation}
\varphi (t) \dot{\varphi}^* (t) - \varphi^*(t) \varphi (t) = i
\end{equation}
guarantees the standard commutation relation
\begin{equation}
[ \hat{a} (t), \hat{a}^{\dagger} (t) ] = 1.
\end{equation}
Hence the time-dependent operators $\hat{a}^{\dagger} (t)$ and
$\hat{a} (t)$ play exactly the same role as the standard creation
and annihilation operators. The Fock space of exact quantum states
consists of the number states of $\hat{a}^{\dagger} (t) \hat{a}
(t)$ \cite{kim-page}
\begin{eqnarray}
\Psi_n (q, t) = \frac{1}{\sqrt{(2 \hbar)^n n!}} \Biggl( \frac{1}{2
\pi \hbar \varphi^*\varphi}\Biggr)^{1/4}
\Biggl(\frac{\varphi}{\sqrt{\varphi^*\varphi}} \Biggr)^{(2n+1)/2}
H_n \Biggl(\frac{q}{\sqrt{2 \hbar \varphi^* \varphi}} \Biggr) \exp
\Biggl[\frac{i}{2 \hbar} \frac{\dot{\varphi}^*}{\varphi^*} q^2
\Biggr], \label{wav fun}
\end{eqnarray}
where $H_n$ is the Hermite polynomial. The density operator
satisfying eq. (\ref{ln eq}) is given by
\begin{equation}
\hat{\rho}_{I} (t) = \frac{1}{Z_{I}} e^{- \hat{I} (t)/ kT_i}, \label{den op}
\end{equation}
where
\begin{equation}
\hat{I} (t) = \hbar \omega_i \Biggl[\hat{a}^{\dagger} (t) \hat{a}
(t) + \frac{1}{2} \Biggr]. \label{quan act}
\end{equation}
Here the free parameters $T_i$ and $\omega_i$ will be fixed as the
temperature and frequency, respectively, if the system starts from
an initial equilibrium.

The action operator, $\hat{I} (t)$, gives an energy analog
$\langle \hat{I} (t) \rangle_{\Psi}$ at each moment, where $\Psi$
denotes the Fock state (\ref{wav fun}), whereas $\langle \hat{H}
(t) \rangle_{\Psi}$ is the dynamical energy. We may write the
expectation value of the density operator as
\begin{eqnarray}
\langle \hat{\rho}_{I} (t) \rangle_{\Psi} = \frac{1}{Z_{I}}
e^{- \langle \hat{I} (t) \rangle_{\Psi}/ kT_i} = \frac{1}{Z_{I}}
e^{- \langle \hat{H} (t) \rangle_{\Psi}/ kT (t)}, \nonumber
\end{eqnarray}
where the time-dependent temperature is given by
\begin{equation}
T (t) \equiv T_i \times \Biggl(\frac{\langle \hat{H} (t) \rangle_{\Psi}}{
\langle \hat{I} (t) \rangle_{\Psi}} \Biggr). \label{tem}
\end{equation}
The process of taking the expectation value with respect to the Fock basis
can be understood from the correspondence principle between quantum
and classical theory. The classical density distribution is
given by
\begin{equation}
\rho_{I} (t) = e^{- I (t)/kT_i}/Z_I,
\end{equation}
where $I(t)$ is the classical action
\begin{equation}
I(t) = \omega_i \Biggl[ a^* (t) a(t) + \frac{1}{2} \Biggr]
\end{equation}
with the classical invariants \cite{kim-page}
\begin{eqnarray}
a^* (t) &=& - i [\varphi (t) p - \dot{\varphi} (t) q] \nonumber\\
a (t) &=& [\varphi^* (t) p - \dot{\varphi}^* (t) q].
\end{eqnarray}
The density distribution also can be written as
\begin{eqnarray}
\rho_{I} (t)
=  \frac{1}{Z_I} \exp\Biggl[- \frac{H(t)}{kT_{c}(t)} \Biggr], \nonumber
\end{eqnarray}
where
\begin{equation}
T_c (t) = T_i \times \Biggl(\frac{H(t)}{I(t)} \Biggr). \label{cl tem}
\end{equation}
The temperature (\ref{tem}) is the quantum analog of the classical
result (\ref{cl tem}), provided that the energy of $H(t)$ and
$I(t)$ is evaluated appropriately in phase space.

A few comments are in order. First, the temperature relation
(\ref{tem}) still holds even when $\vert \Psi \rangle$ is replaced
by $\hat{\rho}_{I} (t)$. From the expectation values
\begin{eqnarray}
\langle \hat{H} (t) \rangle_{{\rho}_I} &=& \frac{\hbar}{2} \coth
\Biggl(\frac{\hbar \omega_i}{2 k T_i} \Biggr) [\dot{\varphi}^* (t)
\dot{\varphi} (t) + \omega^2 (t) \varphi^* (t) \varphi (t)],
\nonumber\\ \langle \hat{I} (t) \rangle_{{\rho}_I} &=& \frac{\hbar
\omega_i}{2} \coth \Biggl(\frac{\hbar \omega_i}{2 k T_i} \Biggr),
\end{eqnarray}
we find the temperature at any later time
\begin{equation}
\frac{T(t)}{T_i} = \frac{\langle \hat{H} (t)
\rangle_{{\rho}_I}}{\langle \hat{I} (t) \rangle_{{\rho}_I}} =
\frac{1}{\omega_i} [\dot{\varphi}^* (t) \dot{\varphi} (t) +
\omega^2 (t) \varphi^* (t) \varphi (t)]. \label{exp tem}
\end{equation}
Second, if the time scale of oscillations determined by $\omega
(t)$ is much smaller than the thermal relaxation time of measuring
device, the time average should be taken over the period:
\begin{equation}
\overline{T} \equiv T_i \times \Biggl(\frac{\overline{\langle \hat{H} (t) \rangle_{\Psi}}}{
\overline{\langle \hat{I} (t) \rangle_{\Psi}}} \Biggr). \label{t-av tem}
\end{equation}
Roughly speaking, the nonequilibrium temperature is determined by
the time-averaged energy of the system at each moment.

\section{Applications}

We now consider the adiabatic or nonadiabatic evolution of boson
oscillator from one static frequency $\omega_i$ to another
$\omega_f$. The solution to eq. (\ref{cl eq}) in general has the
asymptotic form
\begin{eqnarray}
\varphi_i (t) &=& \frac{ e^{- i \omega_i t}}{\sqrt{2 \omega_i}},
\label{ini sol}\\ \varphi_f (t) &=& \alpha \Biggl(\frac{ e^{- i
\omega_f t}}{ \sqrt{2 \omega_f}}\Biggr) + \beta \Biggl(\frac{e^{+
i \omega_f t}}{ \sqrt{2 \omega_f}} \Biggr), \label{fin sol}
\end{eqnarray}
where $\alpha$ and $\beta$ determine the Bogoliubov transformation
(\ref{bog tran}) and satisfy the condition (\ref{bc}). The final
state (\ref{wav fun}), into which eq. (\ref{fin sol}) is
substituted, oscillates with the period $\pi/ \omega_f$. Then the
temperature (\ref{t-av tem}), obtained by taking the time average
of eq. (\ref{exp tem}), is given by
\begin{eqnarray}
\overline{T}_f = T_i \times \Biggl[\frac{\omega_f}{\omega_i} (1 +
2 \beta^* \beta) \Bigg]. \label{tem rel}
\end{eqnarray}
The temperature (\ref{tem rel}) is the same as the temperature
(\ref{nonad tem}) of the nonadiabatic evolution, which was
obtained qualitatively. It also reduces to the adiabatic result
(\ref{app ad tem}) when $|\beta| \ll 1$, that is, the particle
production is negligible. So the temperature (\ref{t-av tem})
recovers the nonadiabatic evolution from one static Hamiltonian in
equilibrium to another.

We investigate the models with a parameter adjusting adiabaticity.
The first model has the frequency which changes according to the
modified P\"{o}schl-Teller type \cite{flu}
\begin{eqnarray}
\omega^2 (t) = \omega_0^2 + \frac{\lambda (\lambda - 1)}{L^2
\cosh^2 (t/\tau)}.
\end{eqnarray}
It has two asymptotic regions at $t \rightarrow \mp \infty$ with
the same frequency $\omega_0$. The strength of interaction is
determined by $\lambda (\lambda -1)$ and the rate and duration by
$\tau$. The factor for pair production is given by
\begin{eqnarray}
\beta^* \beta = \Biggl(\frac{\sin \pi \lambda}{\sinh \pi \omega_0
\tau } \Biggr)^2.
\end{eqnarray}
In the adiabatic limit of large $\tau$  $(\tau \gg 1)$, the
particle production is exponentially suppressed by $e^{- 2 \pi
\omega_0 \tau}$ and the result (\ref{ad tem}) is recovered. One
prominent consequence is the temperature enhancement in the
nonadiabatic limit of small $\tau$ $(\tau \ll 1)$, which leads to
$\beta^* \beta \gg 1$. Another model with an adiabatic parameter
is given by the frequency
\begin{eqnarray}
\omega^2 (t) = \frac{1}{2} (\omega_i^2 + \omega_f^2) - \frac{1}{2}
(\omega_i^2 - \omega_f^2) \tanh \Biggl(\frac{t}{\tau} \Biggr).
\end{eqnarray}
It also has two asymptotic regions at $t = \mp \infty$  with
frequencies $\omega_i$ and $\omega_f$, respectively. It is found
\cite{bir}
\begin{eqnarray}
\beta^* \beta = \frac{\sinh^2 [\pi \tau (\omega_i -
\omega_f)/2]}{\sinh (\pi \tau \omega_i) \sinh(\pi \tau \omega_f)}.
\end{eqnarray}
In the adiabatic limit of large $\tau$, the particle production is
again exponentially suppressed by $e^{- \pi \tau(\omega_i +
\omega_f - |\omega_i - \omega_f |)}$. It is worth noting that in
both models any nonadiabatic process with different $\tau$ leads
to different final temperatures.

The next model is the preheating mechanism of light bosons or
fermions coupled to an oscillating inflaton
\cite{bra1,bra2,kof1,tka,kof2}. To simplify the analysis, we
consider the model frequency of a scalar field mode
\begin{eqnarray}
\omega^2 (t) = \omega_0^2 + \theta (t) \omega_1^2 \cos \omega t,
\end{eqnarray}
where the parametric coupling to the inflaton turns on  at $t =
0$. There are unstable narrow bands, in which the solution to eq.
(\ref{cl eq}) grows exponentially due to the parametric resonance
\begin{eqnarray}
\varphi_f \approx \frac{1}{\sqrt{2 \omega}} \exp \Biggl[- i \omega
t + \frac{\omega_1^2}{2 \omega} t \Biggr].
\end{eqnarray}
Then the temperature (\ref{t-av tem}) is given by
\begin{eqnarray}
\overline{T}_f \approx T_i \times \Biggl[\frac{1}{2} \Biggl\{
\frac{\omega}{\omega_0} + \frac{\omega_0}{\omega} +
\frac{1}{\omega_0 \omega} \Biggl(\frac{\omega_1^2}{2 \omega}
\Biggr)^2 \Biggr\} \Biggr] e^{(\omega_1^2/\omega) t}.
\end{eqnarray}
The thermal energy of exponentially growing temperature comes from
the decaying inflaton. The temperature (\ref{t-av tem}) correctly
describes the growing temperature in the preheating mechanism.

Finally, we extend the above analysis to the nonequilibrium system
of time-dependent fermion oscillator. Pairs of fermions are also
produced when the frequency changes in time \cite{par3,dew}. For
an ensemble of fermion $( \hat{b}, \hat{b}^{\dagger} )$ and its
antipartcle $(\hat{d}, \hat{d}^{\dagger})$ with time-dependent
parameters, pairs of linear invariant operators satisfying Eq.
(\ref{ln eq}) are found in ref. \cite{kim1}:
\begin{eqnarray}
\hat{b} (t) &=& \alpha_{b} (t) \hat{b} + \beta_{b} (t)
\hat{d}^{\dagger}, \quad \hat{b}^{\dagger} (t) = \alpha^*_{b} (t)
\hat{b}^{\dagger} + \beta^*_{b} (t) \hat{d} \\ \hat{d} (t) &=&
\alpha_{d} (t) \hat{d} + \beta_{d} (t) \hat{b}^{\dagger}, \quad
\hat{d}^{\dagger} (t) = \alpha^*_{d} (t) \hat{d}^{\dagger} +
\beta^*_{d} (t) \hat{b}.
\end{eqnarray}
These operators satisfy the anticommutation relations
\begin{eqnarray}
\{ \hat{b} (t), \hat{b}^{\dagger} (t) \} = 1, \quad \{ \hat{d}
(t), \hat{d}^{\dagger} (t) \} = 1,
\end{eqnarray}
and all the other anticommutation relations vanish. The general
result (\ref{tem}) or (\ref{t-av tem}) holds  with the new action
operator
\begin{eqnarray}
\hat{I} (t)  = \hbar \omega_i [ \hat{b}^{\dagger} (t) \hat{b} (t)
- \hat{d} (t) \hat{d}^{\dagger} (t) ].
\end{eqnarray}
So, when the fermion system evolves from one static Hamiltonian
$\hat{H}_i$ with $\omega_i$ in an equilibrium to another
$\hat{H}_f$ with $\omega_f$, we arrive at the final temperature
\begin{eqnarray}
T_f = T_i \times \Biggl[\frac{\omega_f}{\omega_i}( 1 + \beta_{b}^*
\beta_{b} + \beta_{d}^* \beta_{d}) \Biggr]. \label{fer tem}
\end{eqnarray}
The above temperature, which is valid for the adiabatic $(\beta =
0)$ and nonadiabatic $(\beta \neq 0)$ evolution, is the fermion
analog of the boson result (\ref{tem rel}). The coefficient
relations due to the anticommutation relations
\begin{eqnarray}
\alpha_i^* \alpha_j + \beta_i^* \beta_j = \delta_{ij}, \quad i, j
= a, b,
\end{eqnarray}
limit the fermion production due to the Pauli-blocking.

\section{Conclusion and Discussion}

We have studied the nonequilibrium evolution of a time-dependent
boson or fermion system and proposed the nonequilibrium
temperature through eq. (\ref{tem}) or (\ref{t-av tem}). The
quantum statistics is defined quantitatively and rigorously for
the nonequilibrium system under the assumption of the
Schr\"{o}dinger equation (quantum law) and the quantum
Liouville-von Neumann equation. For the time-dependent boson or
fermion oscillator these two equations can be solved
simultaneously in terms of quantum invariant operators, which in
turn can be found in terms of classical solutions of motion. The
true density operator is constructed from a particular quantum
invariant $\hat{I} (t)$, called the action operator. Using the
well-known fact that in the high temperature limit the energy
expectation value of a thermal ensemble of boson or fermion
oscillator is proportional to the temperature, we define the
nonequilibrium temperature (\ref{tem}) or (\ref{t-av tem}) from
the dynamical energy of the Hamiltonian opeartor $\hat{H}(t)$ and
that of the action operator $\hat{I} (t)$ of the true density
operator. It recovers the temperature (\ref{nonad tem}) for the
nonadiabatic evolution from one static Hamiltonian to another, not
necessarily in equilibrium. The nonequilibrium temperature is
applied to many physical models in Sec. VI.

Finally we discuss whether the nonequilibrium temperature
(\ref{tem}) or (\ref{t-av tem}) can be applied to a system whose
final static Hamiltonian has a lower temperature. In the lower
temperature limit the vacuum energy dominates over thermal energy
and leads to the expectation value
\begin{equation}
\langle \hat{H} \rangle_{\hat{\rho}_{H}} = \frac{\hbar \omega}{2}.
\end{equation}
Therefore it seems that the qualitative derivation of temperature
in Sec. II fails in this model. However, the direct analysis below
shows that the temperature (\ref{tem}) or (\ref{t-av tem}) may
still be applied to this model. As a temperature lowering model,
we consider a second order phase transition, which is described by
the sign changing frequency
\begin{equation}
\omega^2 (t) = \theta(- t) \omega_i^2 - \theta (t) \omega_f^2.
\end{equation}
Here $\theta$ is the step function which takes $\theta (t) = 1$
when $t > 0$ and $\theta (t) = 0$ when $t< 0$. The solution to eq.
(\ref{cl eq}) is given by
\begin{eqnarray}
\varphi_i (t) &=& \frac{e^{-i \omega_i t}}{\sqrt{2 \omega_i}},
\nonumber\\ \varphi_f (t) &=& \frac{1}{\sqrt{2 \omega_f}} \Biggl[
\cosh(\omega_f t) - i \frac{\omega_i}{\omega_f} \sinh(\omega_f t)
\Biggr] \nonumber
\end{eqnarray}
The expectation value $\langle \hat{H} (t) \rangle_{\Psi}$ with
respect to any number state (\ref{wav fun}) vanishes because the
kinetic and potential energies contribute equally with opposite
signs \cite{kim-lee}. The final temperature $T_f$ approaches zero
when the instantaneous quench process continues for a sufficiently
long time. In a finite quench the final temperature needs not to
be exactly zero. This implies that the temperature proposed in
this paper may have a wide range of validity including the lower
temperature region.

Not considered in this paper is the process of the nonequilibrium
system evolving toward the final equilibrium state with the true
density operator $\hat{\rho}_{H} = e^{- \hat{H}_f/kT_f}/Z_H$. The
effective theory of time-dependent boson or fermion oscillator
cannot describe this process properly since the final states
oscillate rapidly. To get the non-oscillating final Fock states,
the final solution to eq. (\ref{cl eq}) should have the form
(\ref{ini sol}), which in turn leads to the initial solution of
the form (\ref{fin sol}). This implies that our effective theory
of the open system cannot describe completely the nonequilibrium
process from one equilibrium to another. Even with the more
elaborated model that describes correctly the evolution from one
equilibrium to another, the particle production due to the
nonadiabatic process leads to the chemical potential. These points
together with statistical relations will be addressed in a future
publication.

\vspace{1.5ex}
\begin{flushleft}
{\large\bf Acknowledgements}
\end{flushleft}

The author thanks J.-W. Ho and Prof. W. Israel for stimulating
discussions, which motivated this work, and H.K. Lee, T. Toyoda
and K.H. Yeon for many useful discussions. He also expresses his
appreciation for the warm hospitality of F.C. Khanna and D.N. Page
at the Theoretical Physics Institute, Univ. of Alberta. This work
was supported in part by the Korea Research Foundation under Grant
No. 1998-001-D00364 and 2000-015-DP0080, and also by the Natural
Sciences and Engineering Research Council of Canada.


\begin{thebibliography}{20}

\bibitem{lin} A.D. Linde, Particle Physics and Inflationary
Cosmology (Harwood, Chur, Switzerland, 1990).

\bibitem{bra1} J.H. Traschen and R. H. Brandenberger,
Phys. Rev. D 42  (1990) 2491.

\bibitem{bra2} Y. Shtanov, J. Traschen, and R. Bradenberger,
Phys. Rev. D 51 (1995) 5438.

\bibitem{kof1} L. Kofman, A. Linde, and A.A. Starobinsky, Phys. Rev. Lett. B
76 (1996) 1011.

\bibitem{tka} I. Tkachev, Phys. Lett. B 376 (1996) 35.

\bibitem{kof2} L. Kofman, A. Linde, and A.A. Starobinsky, Phys. Rev. D
56 (1997) 3258.

\bibitem{raj1} K. Rajagoppal and F. Wilczek, Nucl. Phys. B 399 (1993) 395.

\bibitem{raj2}  K. Rajagoppal and F. Wilczek, Nucl. Phys. B 404 (1993) 577.

\bibitem{raj3} K. Rajagoppal and F. Wilczek, The Condensed Matter Physics of
QCD,
hep-ph/0011333.

\bibitem{bun} Y.M. Bunkov and E. Godfrin, Topological
Defects and the Non-Equilibrium Dynamics of Symmetry Breaking
Phase Transitions, Vol. 549 of NATO Advanced Study Institute,
Series C: Mathematical and Physical Sciences (Kluwer Academic,
Dordrecht, 2000).

\bibitem{kim-lee} S.P. Kim and C.H. Lee, Phys. Rev. D 62 (2000) 125020.

\bibitem{d-osc1} H. Dekker, Phys. Rep. 80 (1981) 1.

\bibitem{d-osc2} V.V. Dodonov and V.I. Man'ko, Invariants and
the Evolution of Nonstationary Quantum Systems, edited by M. A.
Markov, Proceedings of the Lebedev Physics Institute, Academy of
Sciences of the USSR, Vol. 183 (Nova Science Pub. Commack, 1989).

\bibitem{d-osc3} C.I. Um, K.H. Yeon, and T. F. George,
The Quantum Damped Harmonic Oscillator, Phys. Rep. (2001) in
press.

\bibitem{kim1} S.P. Kim, A.E. Santana, and F.C. Khanna,
Phys. Lett. A 272 (2000) 46.

\bibitem{par1} L. Parker, Phys. Rev. Lett. 21 (1968) 562.

\bibitem{par2} L. Parker, Phys. Rev. 183 (1969) 1057.

\bibitem{dew} B.S. DeWitt, Phys. Rep. 19C (1975) 297.

\bibitem{lew} H.R. Lewis, Jr. and W.B. Riesenfeld, J. Math. Phys. (N.Y.)
10 (1969) 1458.

\bibitem{kim2} J.K. Kim and S.P. Kim, J. Phys. A 32 (1999) 2711.

\bibitem{kim-page} S.P. Kim and D.N. Page, Phys. Rev. A 64 (2001)
012104.

\bibitem{flu} S. Fl\"{u}gge, Practical Quantum Mechanics
(Springer-Verlag, Berlin, 1974).

\bibitem{bir} N. D. Birrel and P. C. W. Davies,
Quantum Fields in Curved Space (Cambridge Univ. Press, Cambridge,
1982).

\bibitem{par3} L. Parker, Phys. Rev. D 3 (1971) 346 (1971).


\end{thebibliography}
\end{document}